\documentclass[
 reprint,
superscriptaddress,
 amsmath,amssymb,
prl,
floatfix,
]{revtex4-2}

\usepackage{graphicx}
\usepackage{dcolumn}
\usepackage{bm}
\usepackage{color}
\usepackage{float}
\usepackage[colorlinks,urlcolor=blue,citecolor=blue,linkcolor=blue]{hyperref}

\begin{document}

\preprint{topo}

\title{Motional narrowing, ballistic transport, and trapping of room-temperature exciton polaritons in an atomically-thin semiconductor}

\author{M.~Wurdack}
 \affiliation{ARC Centre of Excellence in Future Low-Energy Electronics Technologies and Nonlinear Physics Centre, Research School of Physics, The Australian National University, Canberra, ACT 2601, Australia}

\author{E.~Estrecho}%
 \affiliation{ARC Centre of Excellence in Future Low-Energy Electronics Technologies and Nonlinear Physics Centre, Research School of Physics, The Australian National University, Canberra, ACT 2601, Australia}

\author{S.~Todd}
 \affiliation{ARC Centre of Excellence in Future Low-Energy Electronics Technologies and Nonlinear Physics Centre, Research School of Physics, The Australian National University, Canberra, ACT 2601, Australia}

\author{T.~Yun}%
 \affiliation{ARC Centre of Excellence in Future Low-Energy Electronics Technologies and Nonlinear Physics Centre, Research School of Physics, The Australian National University, Canberra, ACT 2601, Australia}

\author{M.~Pieczarka}
 \affiliation{ARC Centre of Excellence in Future Low-Energy Electronics Technologies and Nonlinear Physics Centre, Research School of Physics, The Australian National University, Canberra, ACT 2601, Australia}
\affiliation{Department of Experimental Physics, Wroc\l{}aw University of Science and Technology, Wyb.~Wyspia\'nskiego 27, 50-370 Wroc\l{}aw, Poland}%

\author{S.~K.~Earl}
\affiliation{ARC Centre of Excellence in Future Low-Energy Electronics Technologies and Centre for Quantum and Optical Science, Swinburne University of Technology, Victoria 3122, Australia}%

\author{J.~A.~Davis}
\affiliation{ARC Centre of Excellence in Future Low-Energy Electronics Technologies and Centre for Quantum and Optical Science, Swinburne University of Technology, Victoria 3122, Australia}%

\author{C.~Schneider}
\affiliation{Institut f\"ur Physik, Carl von Ossietzky Universit\"at Oldenburg, Ammerl\"ander Heerstra{\ss}e 114-118, 26126 Oldenburg, Germany}%

\author{A.~G.~Truscott}%
 \affiliation{Laser Physics Centre, Research School of Physics, The Australian National University, Canberra, ACT 2601, Australia}
 
 \author{E.~A.~Ostrovskaya}
 \affiliation{ARC Centre of Excellence in Future Low-Energy Electronics Technologies and Nonlinear Physics Centre, Research School of Physics, The Australian National University, Canberra, ACT 2601, Australia}

\maketitle

\noindent \textbf{Atomically-thin transition metal dichalcogenide crystals (TMDCs) hold great promise for future semiconductor optoelectronics \cite{TMDC2012} due to their unique electronic and optical properties. In particular, electron-hole pairs (excitons) in TMDCs are stable at room temperature and interact strongly with light \cite{Schneider2018}. When TMDCs are embedded in an optical microcavity, the excitons can hybridise with cavity photons to form exciton polaritons (polaritons herein) \cite{Liu2015}, which display both ultrafast velocities and strong interactions. The ability to manipulate and trap polaritons on a microchip is critical for future applications \cite{Sanvitto2016,Schneider2016}. Here, we create a potential landscape for room-temperature polaritons in monolayer WS$_2$, and demonstrate their free propagation and trapping. We show that the effect of dielectric disorder, which restricts the diffusion of WS$_2$ excitons \cite{Zipfel2020} and broadens their spectral resonance \cite{Raja2019}, is dramatically reduced in the strong exciton-photon coupling regime leading to motional narrowing \cite{Whittaker1996}. This enables the ballistic transport of WS$_2$ polaritons across tens of micrometers with an extended range of partial first-order coherence. Moreover, the dephasing of trapped polaritons is dramatically suppressed compared to both WS$_2$ excitons and free polaritons. Our results demonstrate the possibility of long-range transport and efficient trapping of TMDC polaritons in ambient conditions.}

Polaritons are bosonic quasi-particles consisting of bound excitons and confined photons, which form in optical microcavities with embedded direct bandgap semiconductors \cite{Microcavities,Weissbuch1992} in the strong exciton-photon coupling regime. They inherit large group velocities from their photonic component, and interact due to their excitonic component, which enables them to display collective quantum phenomena \cite{Kasprzak2006,Deng2010,Byrnes2014,Amo2009} in a solid state. A roadmap of polariton-based optoelectronic devices \cite{Sanvitto2016} suggests multiple applications, including ultra-low threshold lasers \cite{Schneider2012} and novel computing architectures \cite{Ballarini2013}. Proof-of principle demonstrations of these applications often rely on potential landscape engineering for the polaritons, e.g, by lithographic patterning, which is a well-established, advanced technology for epitaxially grown GaAs-based microcavities \cite{Schneider2016}. However, due to the low exciton binding energies in III-V semiconductor systems, their operation is limited to cryogenic temperatures. Although polariton condensation and trapping in engineered potential landscapes at room temperature were demonstrated by utilising semiconductors with large exciton binding energies \cite{Plumhof2013,Daskalakis2014,ZnO2015,Lerario2017,Su2018,Su2020,Dusel2020}, the search for the optimal polaritonic material platforms that combine stability of the samples and low disorder continues \cite{Sanvitto2016}.

\begin{figure}[ht!]
\centering
\includegraphics[width=8.6cm]{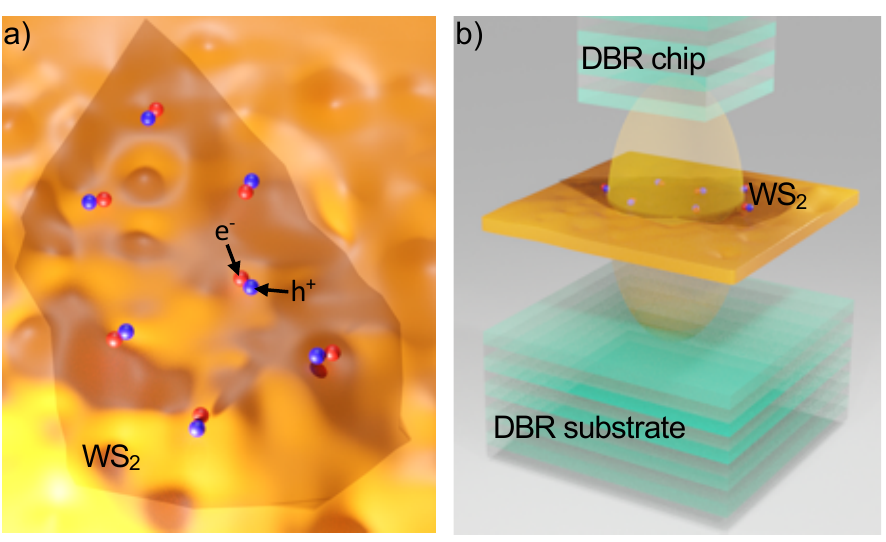}
\caption{Schematics of {\bf a)} monolayer WS$_2$ hosting bound electron-hole pairs (excitons) placed on a substrate with substantial dielectric disorder with the spatial scale comparable with the exciton size \cite{Zipfel2020}. The electrons and holes are represented by red (${\rm e}^-$) and blue (${\rm h}^+$) balls, respectively.  {\bf b)} Hybridisation of excitons and photons in an all-dielectric high-Q optical microcavity reduces the effect of dielectric disorder \cite{Whittaker1996}. }
\label{fig:Fig1}
\end{figure}

Recently, atomically-thin, two-dimensional (2D) crystals of transition metal dichalcogenides (TMDCs) have emerged as extremely promising candidates for room-temperature polaritonics due to the large exciton binding energies and strong light-matter interactions \cite{Schneider2018}. Striking properties of polaritons \cite{Liu2015}, such as the spin-valley Hall effect \cite{Lundt2019}, the formation of electrically charged polaron polaritons \cite{Sidler2017}, and signatures of bosonic condensation \cite{Waldherr2018,Carlos2020,Zhao2020} were explored in these systems. 

Here, we demonstrate room-temperature WS$_2$ polaritons under non-resonant continuous-wave (cw) optical excitation in a high-quality all-dielectric monolithic microcavity with a non-trivial potential landscape. This potential landscape allows us to investigate both freely moving and trapped WS$_2$ polaritons in the ``thermal" regime, below the onset of bosonic condensation. Excitons in monolayer WS$_2$ were previously shown to be profoundly affected by dielectric disorder in the environment (Fig.~\ref{fig:Fig1}a), which causes inhomogeneous linewidth broadening \cite{Raja2019} and reduction of the exciton diffusion coefficient due to scattering \cite{Zipfel2020}. By analysing the spectrum and the first order coherence of WS$_2$ excitons strongly coupled to the cavity photons, we find dramatic motional narrowing \cite{Whittaker1996} of the inhomogeneously broadened linewidth and suppression of disorder-induced dephasing (Fig.~\ref{fig:Fig1}b) at room temperature. Due to the drastic suppression of disorder-induced effects, WS$_2$ polaritons exhibit strong signatures of ballistic, dissipationless propagation with conservation of total (potential and kinetic) energy and partial coherence. This enables polariton trapping in a quasi-1D potential well, even when the excitation spot is located tens of micrometers away from the trap.  

The all-dielectric monolithic microcavity investigated in this work was fabricated with the flip-chip approach \cite{Lundt2019,Rupprecht2021}, in which a small piece of a dielectric Bragg reflector (DBR) was transferred from a polypropylene carbonate (PPC) film onto a DBR substrate, at the position of a mechanically exfoliated WS$_2$ monolayer (see Methods and Supplementary Information for details). This fabrication process fully maintains the excitonic properties of monolayer WS$_2$ since it does not involve any direct material deposition on top of the monolayer, which can cause strong exciton quenching \cite{Kim2018,Wurdack2021}.

\begin{figure}[ht!]
\centering
\includegraphics[width=8.6cm]{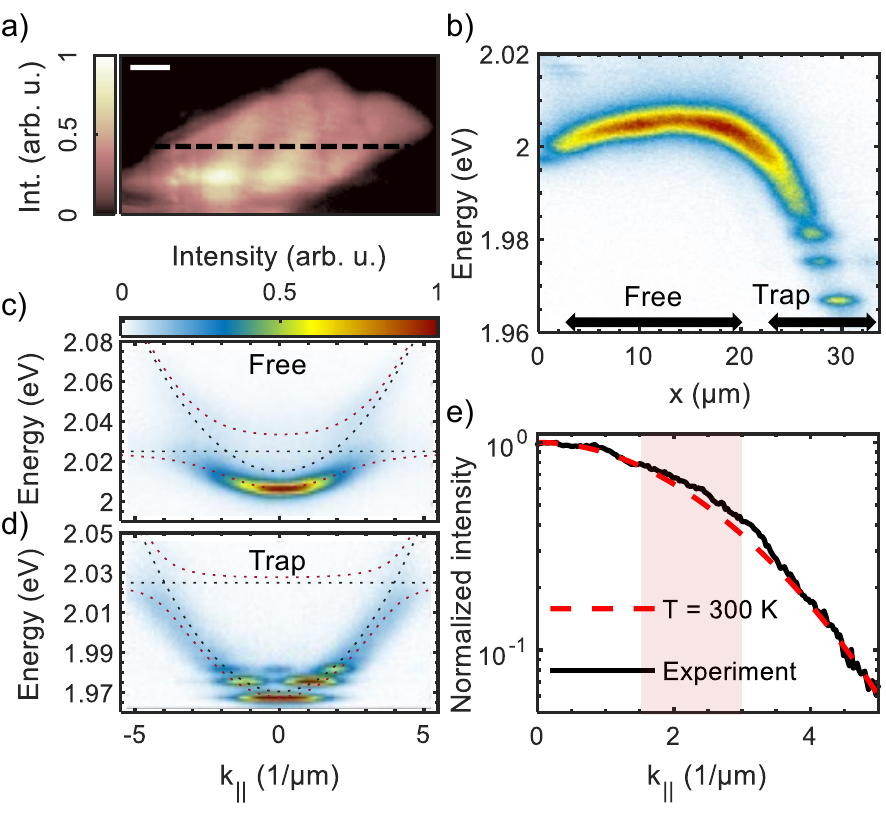}
\caption{{\bf a)} PL map of the monolayer region inside the microcavity. The scale bar corresponds to $5~\mathrm{\mu m}$. {\bf b)} Potential energy of the polaritons across the dashed line in  panel (a) extracted from the PL signal at $k_{||}\approx0$. {\bf c,d)} Polariton dispersion extracted from the angle-resolved photoluminescence spectra of the sample (c) in the region with a weak potential gradient (free) and (d) in the trap region (trap). The flat (parabolic) black dotted lines correspond to the exciton(cavity photon) energies and the upper(lower) red dotted lines are the fitted upper(lower) polariton branches with the Rabi splitting $2\hbar \Omega\approx25~\mathrm{meV}$. {\bf e)} Intensity profile of the PL spectrum in (c) with the theoretical intensity profile of a fully thermalised polariton gas at $T=300$ K. The bottleneck region of the free polariton dispersion (between $k\approx 1.5~\mu$m$^{-1}$ and $k\approx 3~\mu$m$^{-1}$) is shaded.}
\label{fig:Fig2}
\end{figure}

When the microcavity is excited by a continuous wave (cw) frequency doubled ND:YAG laser through a DBR Bragg minimum at $\lambda = 532~\mathrm{nm}$, the intensity map of the photoluminescence (PL) at the position of the monolayer shows strong polariton emission (see Fig.~\ref{fig:Fig2}a). Measuring the polariton emission spectrum along the dashed line at the angles of approximately zero incidence (see Methods) allows us to estimate the profile of the potential landscape \cite{Pieczarka2019} for polaritons corresponding to zero kinetic energy ($k_{||}\approx0$). This measurement reveals the non-trivial shape (see Fig.~\ref{fig:Fig2}b) of the potential caused by strong variation of the detuning between the cavity photon energy, $E_C$, and the exciton resonance, $E_X$. This variation is caused by an air gap between the DBR chip and the DBR substrate, which leads to a local modification of the cavity length. 

In contrast to the approximately linear cavity wedge in GaAs-based microcavities \cite{Snoke2013}, the variation of the cavity length (and $E_C$) in our system is highly nonlinear.  As a result, the polariton ground state (i.e., its energy at $k_{||}=0$) considerably varies with position at $x>20~\mathrm{\mu m}$ and an effective trap for polaritons is formed at $x=30~\mathrm{\mu m}$. The trap is a quasi-1D potential well strongly elongated in the direction perpendicular to the measurement direction, $x$ (see Supplementary Information). The angle resolved PL spectra (dispersion) in the ``planar" region with a weak potential gradient ($x<20~\mathrm{\mu m}$) (Fig. \ref{fig:Fig2}c) and in the effective trap (Fig. \ref{fig:Fig2}d) can be well fitted with the lower (LP) and upper (UP) polariton branches $E_{LP/UP} = \frac{1}{2}[E_X + E_C\pm \sqrt{(2\hbar\Omega)^2 + \delta^2}]$, where $\delta=E_X - E_C$ is the exciton-photon energy detuning, and $2\hbar\Omega$ is the Rabi splitting. Therefore, the sample operates in the strong exciton-photon coupling regime at room temperature across its whole area, within a large range of detunings $\delta \in [10, 60]~\mathrm{meV}$. The detuning defines the exciton fraction of the polariton through the excitonic Hopfield coefficient \cite{Deng2010}, which takes values $|X^F|^2 \approx 0.3$ and $|X^T|^2 \approx 0.05$ for the free and trapped polaritons, respectively (see Supplementary Information).

The occupation numbers of the momentum states of the free polaritons (at $x=16~\mathrm{\mu m}$), with the dispersion shown in Fig.~\ref{fig:Fig2}c,  are reflected in the intensity of the PL profile and can be well described with a model of a fully thermalised polariton gas \cite{Lundt2016} (see Supplementary Information) at room temperature $T=300~\mathrm{K}$ (see Fig.~\ref{fig:Fig2}e). In the well established model for the polariton energy relaxation under non-resonant excitation \cite{Byrnes2014}, polaritons reduce their kinetic energy by scattering with excitons, other polaritons and phonons. While the effective inter-particle interactions in our structure are weak due to the large binding energies and small Bohr radii of excitons \cite{Shahnazaryan2017}, the observed thermalisation implies thermal equilibrium of the polariton gas with the environment and efficient polariton-phonon interactions at room temperature.  The slight departure towards the higher occupation numbers in Fig.~\ref{fig:Fig2}e occurs around the inflection point of the polariton dispersion (see Supplementary Information), referred to as relaxation bottleneck \cite{Byrnes2014}, where the effective mass changes sign and interaction with phonons becomes less efficient.

Quantisation of the trapped polariton spectrum is highly pronounced in Fig. \ref{fig:Fig2}d because the lateral size of the trap in $x$-direction (a few micrometers) is comparable to the thermal de Broglie wavelength of polaritons \cite{Baumberg2018}, $\lambda_{\rm th}= \sqrt{2 \pi \hbar^2\left(m_{\rm eff} k_B T\right)^{-1}}$ \cite{Damm2017}, where $m_{\rm eff}$ is the polariton effective mass extracted from the dispersion fits in Fig. \ref{fig:Fig2}(c,d). At $T=300$ K, we obtain for the free and trapped polaritons $\lambda_{\rm th}^F  = (1.01\pm 0.05) ~\mathrm{\mu m}$ and  $\lambda_{\rm th}^T  = (1.18\pm 0.05) ~\mathrm{\mu m}$, respectively. The ground, first, and second excited states in the trap are clearly occupied, and the energies of these states fit well to the simulation results (see Supplementary Information). 

\begin{figure}
\centering
\includegraphics[width=8.6cm]{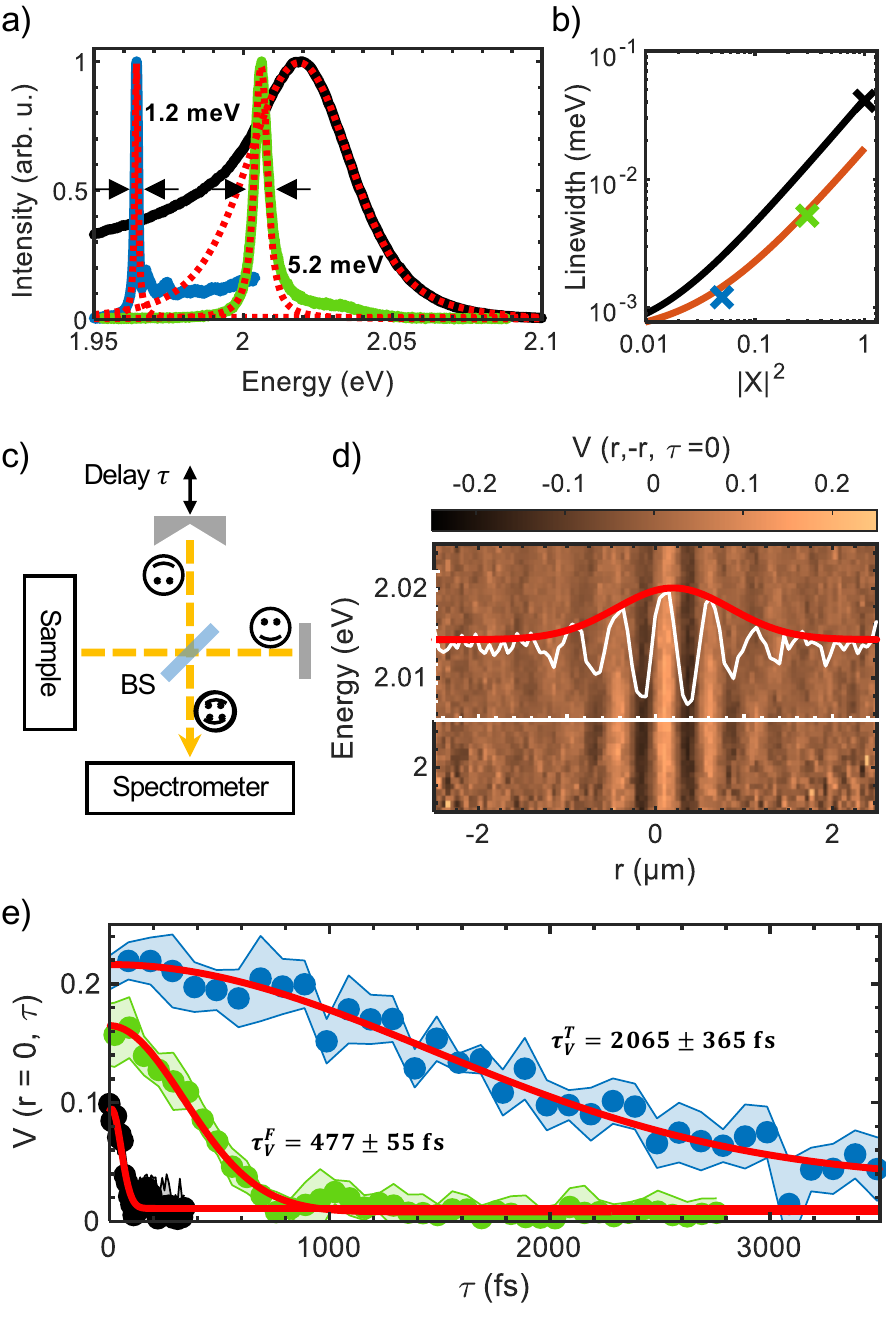}
\caption{{\bf a)} PL spectra at k$_{||} = 0$ of the free polaritons (green), trapped polaritons (blue), and of a WS$_2$ exciton in the bare monolayer on the same substrate (black). The WS$_2$ exciton spectrum and the polariton ground states were fitted with a Voigt line profile (red dots). {\bf b)} Theoretical polariton linewidth as a function of excitonic Hopfield coefficient, based on the WS$_2$ exciton linewidth (see text) and a cavity mode with a quality factor $Q\approx3000$, with (black) and without (orange) contribution of inhomogeneous exciton linewidth broadening. Crosses mark the linewidths of the exciton (black) and polariton (green and blue) emissions from the data in panel (a). {\bf c)} Schematic of the experimental setup of the modified Michelson interferometer (see Methods).  {\bf d)} Normalised, spectrally resolved interferogram of the free polaritons at zero time delay (see Methods) superimposed with the  interference fringes at their lowest energy $E=2.004~\mathrm{eV}$ in the range $[-0.2, 0.2]$ (white) and the corresponding fringe visibility (red). {\bf e)} Time coherence measurements of the excitons (black data points), free polaritons (green data points) and trapped polaritons (blue data points) at their lowest energies, fitted with a second order exponential decay function (red lines). The extracted decay times for the fringe visibility are $\tau_V^X=(62\pm15)~\mathrm{fs}$ for the  excitons, $\tau_V^F=(477\pm55)~\mathrm{fs}$ for the free polaritons, and $\tau_V^T=(2065\pm365)~\mathrm{fs}$ for the trapped polaritons. The shaded areas correspond to the uncertainties of the fitting procedure and represent a 95\% confidence interval.}
\label{fig:Fig3}
\end{figure}

To quantify the effect of the dielectric disorder on the WS$_2$ polaritons at room temperature, we investigate the PL of both free and trapped polaritons at $k_{||}=0$ by employing filtering in $k$-space (see Methods), and compare it to the PL of bare excitons in a WS$_2$ monolayer on the same DBR substrate. The PL spectra presented in Fig. \ref{fig:Fig3}a are fitted with convoluted Lorentzian and Gaussian distributions (Voigt function), corresponding to the homogeneous and inhomogeneous linewidth broadening, respectively. The homogeneous broadening is mainly determined by the radiative decay of the excitons and exciton-phonon interactions \cite{Selig2016}, and the inhomogeneous broadening is largely due to dielectric disorder on the substrate surface, which causes local fluctuations of the exciton binding energies \cite{Raja2019}. For the excitons, we extract a total linewidth of $\Delta E_X=(41.5\pm0.5) ~\mathrm{meV}$, with the homogeneous linewidth broadening of $\Delta E_X^{H}=(17.5 \pm 0.3)~\mathrm{meV}$ and the inhomogeneous linewidth broadening of $\Delta E_X^{IH} =(31.3 \pm 0.2)~\mathrm{meV}$ within $95$\% confidence level. The homogeneous linewidth broadening corresponds to $\gamma^{-1}_X=2\hbar/ \Delta E_{X}^{H}=\left(75 \pm 2\right)~\mathrm{fs}$, which is in good agreement with the decoherence rate of excitons in a WS$_2$ monolayer placed on a high-quality SiO$_2$ substrate, as determined in the four-wave mixing experiments (see Supplementary Information).

Compared to the excitons, both free and trapped polaritons display much narrower linewidths (see Fig. \ref{fig:Fig3}a) of $\Delta E_F=\left(5.2\pm0.1\right) ~\mathrm{meV}$ and $\Delta E_T=\left(1.20\pm0.01\right)~\mathrm{meV}$, respectively. The inhomogeneous broadening is $\Delta E_F^{IH}=(2.6 \pm 0.1)~\mathrm{meV}$ for the free polaritons and is negligible for the trapped polaritons. Moreover, the linewidths of the free and trapped lower polaritons are substantially smaller than the theoretically calculated (see Methods) polariton linewidth (Fig.~\ref{fig:Fig3}b, black line). The theoretical and the measured linewidths agree well only when the inhomogeneous exciton broadening is completely eliminated from the calculation (Fig.~\ref{fig:Fig3}b, orange line). This indicates that the effects causing linewidth broadening are significantly reduced for the WS$_2$ excitons strongly coupled to the cavity photons. This so-called motional narrowing is a well-known effect in quantum well microcavities \cite{Whittaker1996, Houdre1996, Savona1997}, and was recently observed at cryogenic temperatures for excitons in monolayer MoSe$_2$ strongly coupled to cavity photons \cite{Dufferwiel2015} and to optical bound states in the continuum \cite{Kravtsov2020}. Although the exciton-photon coupling strength in our system ($2\hbar\Omega$) does not exceed the disorder-induced fluctuations of exciton binding energy (quantified by $\Delta E_X^{IH}$) \cite{Houdre1996}, the motional narrowing still occurs due to the size of polaritons, $\lambda_{\rm th}$, significantly exceeding the spatial scale of these fluctuations \cite{Whittaker1996}, with the magnitude of the effect exceeding that observed in quantum well microcavities by an order of magnitude.

In addition to the linewidth broadening, dielectric disorder causes rapid dephasing of WS$_2$ excitons (see Supplementary Information). To compare the timescales of dephasing between the excitons and polaritons, we perform the coherence measurements with a modified Michelson interferometer (see Fig.~\ref{fig:Fig3}c) \cite{Kasprzak2006}. In this configuration, overlapping the original image $I_{o}$ with its flipped and time-delayed copy $I_{f}$ causes interference fringes with the visibility (see Methods):
\begin{equation}
V(r, -r, \tau) = \left[I_{t} - \left(I_{o} + I_{f}\right)\right]\left(2\sqrt{I_{o} I_{f}}\right)^{-1}, \nonumber
\end{equation}
where $r$ is the distance from the axis of the retro-reflector, $\tau$ the time delay between the interferometer arms, and $I_{t}$ is the total intensity at $r$ and $\tau$ (see Methods). The envelope of $V(r, -r, \tau)$ is a measure for the absolute value of the first order coherence function $\left|g^{(1)}(r, \tau)\right|$, and quantifies dephasing between $I_{o}$ and $I_{f}$. For the coherence measurement, we remove the $k$-space filtering to avoid filter artefacts and employ a spectrometer to be able to compare the dephasing times of the excitons, free polaritons and trapped polariton in their lowest energy states. The normalised, spectrally resolved interference pattern of the free polaritons at zero-delay $\tau = 0$ is shown in Fig. \ref{fig:Fig3}d, together with $V(r, -r, \tau=0)$ and its envelope as a measure for $\left|g^{(1)}(r, \tau=0)\right|$. The full-width at half maximum ${\rm FWHM}(\left|g^{(1)}\right|) = \left(1.5 \pm 0.1\right)~ \mathrm{\mu m}$ is much larger than the theoretically expected value \cite{Damm2017} for the thermal polariton gas $\lambda_{\rm th}\sqrt{\ln 2/(4\pi)}\approx 0.24 ~\mathrm{\mu m}$. This indicates that, as a result of diminished effects of dielectric disorder, the polaritons expand without scattering-induced dephasing.

The time coherence measurements are performed at $\tau\neq0$. Figure \ref{fig:Fig3}e shows the envelopes of $V(r=0,\tau)$ for the excitons, free polaritons, and trapped polaritons around their lowest energies, fitted with a second order exponential decay function \cite{Reimer2016}. We define the time at which $V(r=0,\tau)$ drops to $1/\mathrm{e}$ as the decay time of the visibility $\tau_V$. While the absolute values of $\tau_V$ extracted from the data in Fig.~\ref{fig:Fig3}e are most likely larger than the actual dephasing times due to the spatial resolution of our experimental setup \cite{Damm2017}, their relative values scale with the relative linewidths of the excitons, free polaritons, and trapped polaritons as: $\Delta E_X/\Delta E_F = \tau_V^F/\tau_V^X \approx 8$, and $\Delta E_X/\Delta E_T = \tau_V^T/\tau_V^X \approx 33$, showing that the changes in linewidths directly correlate with the changes in dephasing times. Hence, due to the motional narrowing the dephasing of WS$_2$ polaritons is not affected by scattering on dielectric disorder and is exclusively determined by their intrinsic decoherence and lifetimes. 

\begin{figure}
\includegraphics[width=8.6cm]{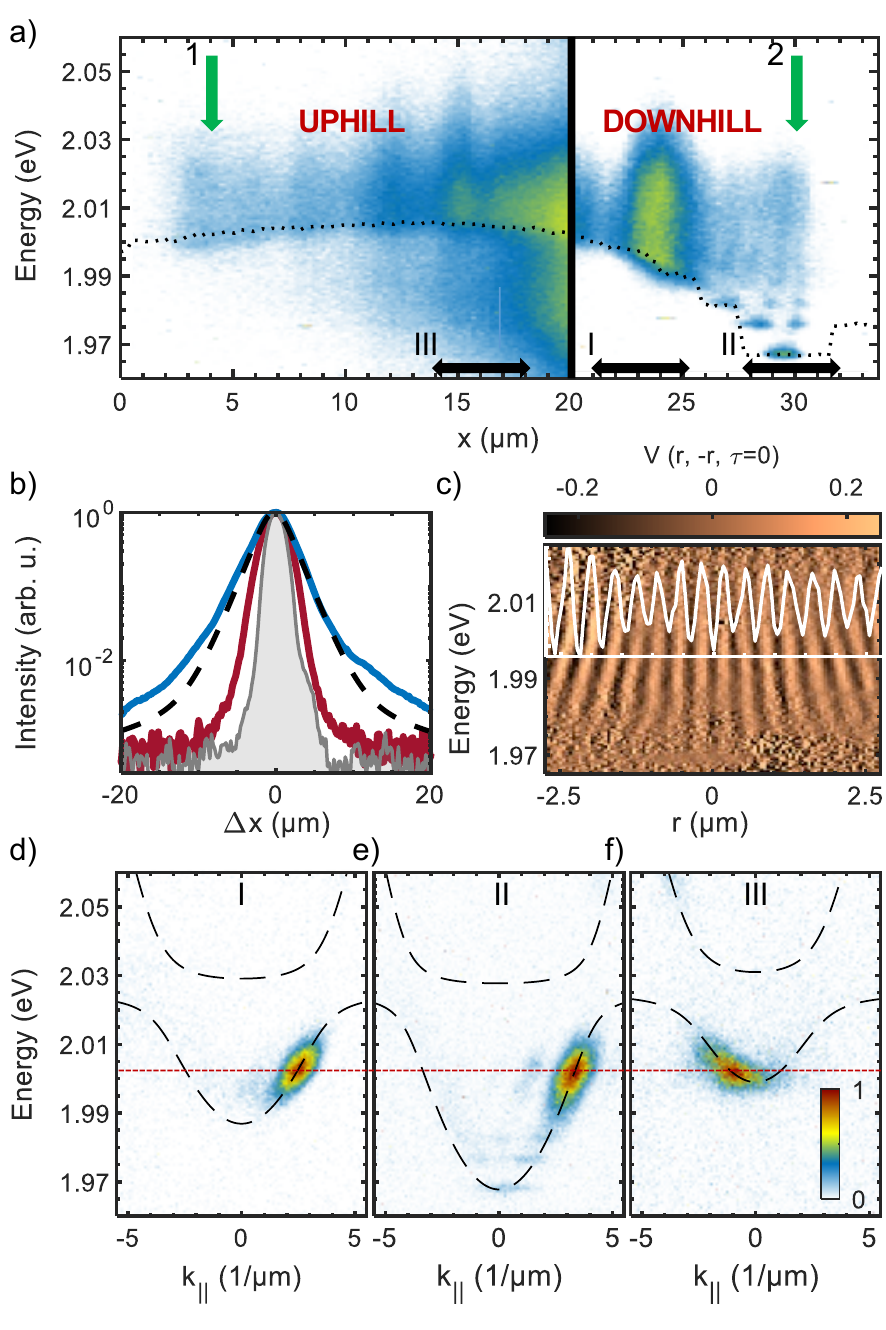}
\caption{{\bf a)} Combined position-resolved PL spectrum of the polaritons propagating in the potential landscape (left panel) uphill from the excitation spot (green arrow) at position (2), $x=30~\mathrm{\mu m}$, and (right panel) downhill from the excitation spot (green arrow) at position (1) $x=4~\mathrm{\mu m}$. The outline of the potential landscape obtained by extracting spectral peak position from the $k_{||}\approx0$ signal in Fig. \ref{fig:Fig2}b is marked with a dotted black line. The colormap is logarithmic. {\bf b)} Real-space profile of (gray) the laser excitation spot, and the corresponding PL profiles of (red) exciton emission of a monolayer WS$_2$ on the DBR substrate, and (blue) polariton emission of the microcavity. The excitation spot was positioned at $x=16~ \mathrm{\mu m}$. The black dashed line is the modelled PL profile of the thermal polariton gas at $T=300~\mathrm{K}$ (see Supplementary Information). {\bf c)} Normalized, spectrally resolved interferogram of the travelling polaritons at $x=30~\mathrm{\mu m}$, when excited at $x= 23~\mathrm{\mu m}$. The (white) normalized interference fringes at $E=1.995~\mathrm{eV}$ in the range $[-0.2, 0.2]$ are plotted on top. {\bf d,e)} Normalised angle-resolved PL spectra of regions (d) I and (e) II marked in panel (a), when excited at $x= 16~\mathrm{\mu m}$, and {\bf f)} angle-resolved PL spectra of region III in panel (a) when excited at $x= 30~\mathrm{\mu m}$. The spectra in (d-f) were measured with different spectrometer settings and aquisition times and then normalised to compare the positions of the travelling wave packets in the ($k_{||}$,E) parameter space rather than their intensities. The red dashed line marks $E=2.003~\mathrm{eV}$}
\label{fig:Fig4}
\end{figure}

Scattering of bare WS$_2$ excitons on disorder also causes energy dissipation and strong reduction of the diffusion coefficient  \cite{Zipfel2020}. In order to contrast the propagation of the WS$_2$ polaritons with excitons, we collect the PL at the positions next to the excitation spot (see Methods). Figure~\ref{fig:Fig4}a shows the real-space resolved PL spectra along the dashed line in Fig.~\ref{fig:Fig2}a, when exciting the sample either at $x=4~\mathrm{\mu m}$ (position 1) and collecting the PL at positions $x>20~\mathrm{\mu m}$ (downhill), or at $x=30~\mathrm{\mu m}$ (position 2) and collecting the PL at positions $x<20~\mathrm{\mu m}$ (uphill). Strikingly, regardless of the excitation position, the polaritons travel across the whole monolayer area, both up and down the potential gradient shown in Fig.~\ref{fig:Fig2}b, which illustrates the long-range propagation of the WS$_2$ polaritons exceeding the mean exciton diffusion length of WS$_2$ excitons (360 nm \cite{Kulig2018}) by orders of magnitude.

By measuring the real-space PL profile when exciting either the microcavity or the bare monolayer WS$_2$ with a focused laser spot (positioned at $x=16~\mathrm{\mu m}$ for the microcavity sample), we can quantitatively compare the WS$_2$ polariton transport with the diffusion of WS$_2$ exciton. Figure~\ref{fig:Fig4}b shows the emission profiles of the microcavity polaritons, the monolayer excitons, and the laser spot on logarithmic scales. It is clear that the line shape of the polariton PL strongly differs from the approximately Gaussian line profile of the laser and the monolayer PL. This is caused by the Boltzmann distribution of the thermalised polariton gas \cite{Lundt2016}, in which the upper polariton branch and the lower polariton branch at higher $k_{||}$ values are well occupied at room temperature (see Fig.~\ref{fig:Fig2}e). The corresponding modelled PL profile of the polariton gas at $T=300~\mathrm{K}$ expanding with the maximum group velocity (black dashed line in Fig. \ref{fig:Fig4}b) qualitatively reproduces the measured profile, and calculations for different temperatures show that the rapid expansion of the polariton gas from the pump spot is promoted by its high temperature (see Supplementary Information). The extend of the measured profile is larger than the calculated profile most likely due to the higher polariton occupation numbers in the bottleneck region compared to the fully thermalised polariton gas (see Fig.~\ref{fig:Fig2}b). While the exciton PL intensity drops to $1$\% at $\Delta x\approx 5~\mathrm{\mu m}$, the polariton intensity reaches this level at $\Delta x\approx 12~\mathrm{\mu m}$. For comparison, the laser excitation drops to this intensity level at $\Delta x\approx 3~\mathrm{\mu m}$, and with a (sub)linear relationship between excitation intensity and PL intensity \cite{Hoshi2017} the transport length of the polaritons in this structure is at least $5$ times larger compared to the excitons.

The spatial coherence of the travelling polaritons was investigated by exciting the structure on position I in Fig.~\ref{fig:Fig4}a and measuring the interferogram of the travelling wave packet at position II (see Fig.~\ref{fig:Fig4}c), i.e., in the trap region. The interference fringes, extracted by removing the Doppler effect, have a substantial spatial extent, and their visibility is similar to the maximum visibility of the interference pattern at the excitation spot (see Fig. \ref{fig:Fig3}d). The extent of the spatial coherence can be well explained by the high kinetic energy of the travelling wave packet, and the conservation of the magnitude of the fringe visibility is a direct result of low dephasing. This finding further confirms that scattering with the disorder in the dielectric environment, which causes dephasing, is heavily reduced in the strongly light-matter coupled system.

Finally, we analyse the dispersion of the propagating polaritons by collecting the angle-resolved PL spectra from small areas in the real space (see Methods). In contrast to real-space resolved spectral imaging (Fig.~\ref{fig:Fig4}a), in which the emitted PL is collected in all directions of the light cone, angle-resolved spectral imaging collects the PL signal limited to angles of incidences along the spectrometer slit direction. This allows us to characterise the energy spectrum of the polaritons in one particular direction $x$, i.e. along the dashed line in Fig.~\ref{fig:Fig2}a. First, we measure PL at positions I and II in Fig~\ref{fig:Fig4}a, when exciting the sample at $x=16~\mathrm{\mu m}$ (position III). The spectra (Fig.~\ref{fig:Fig4}d,e) unveil that the polaritons are travelling at a constant energy centred at $E \approx 2.003~\mathrm{eV}$ (red dashed line in Fig.~\ref{fig:Fig4}d,e) along the potential gradient, and that their potential energy (with respect to the global minimum) is almost fully converted into kinetic energy (i.e., the energy at $k$ measured with respect to the energy at $k=0$). The constant energy of the propagating polaritons approximately coincides with the inflection point of the polariton dispersion, i.e. the maximum value of its group velocity (see Supplementary Information). This apparent lack of the energy relaxation along the gradient is due to the low effective inter-particle interactions \cite{Shahnazaryan2017}, and suppression of disorder-induced scattering, which results in the reduced energy dissipation in the system. Hence, room-temperature polaritons in WS$_2$ can propagate ballistically over at least tens of micrometers.

Remarkably, when swapping the excitation and detection positions, i.e. exciting the polaritons at the position of the trap (position 2 in Fig.~\ref{fig:Fig4}a), and measuring the angle-resolved PL spectra at position III in Fig.~\ref{fig:Fig4}a, we find that the energy of the polaritons moving up the potential hill is approximately the same as the energy of the downhill flow, see Fig.~\ref{fig:Fig4}f. This effect is also detectable in the upper polariton branch, as discussed in Supplementary Information. This uphill flow is due to the high-momenta thermalised polaritons excited in the trap region (Figure~\ref{fig:Fig2}d) with the above-barrier kinetic energies. Without energy dissipation, these polaritons efficiently convert their high kinetic energy into potential energy while flowing uphill and populate the planar region of the sample, as observed in Fig.~\ref{fig:Fig4}a.

Despite the constant energy flow along the gradient ($x$-direction), clear energy relaxation and the resulting occupation of the low-energy trapped states is visible in the position-resolved spectral image (Fig. \ref{fig:Fig4}a, region II), which collects polariton emission from all directions, including that orthogonal to the quasi-1D trap. Weak signatures of this relaxation are visible in Fig. \ref{fig:Fig4}e, but the signal is stronger for the emission not filtered along $x$. This indicates that  the phonon-induced energy relaxation for room temperature polaritons is sufficient to drive the occupation of the lower energy states in the trap. The trap is occupied by polaritons even with the excitation spot located tens of micrometers away at the opposite side of the monolayer, i.e. at position (1) in Fig. \ref{fig:Fig4}a. 

In summary, we have realised freely moving and trapped WS$_2$ polaritons in a non-trivial potential landscape at room temperature. The pronounced motional narrowing and suppressed dephasing of the polaritons point to dramatic reduction of the effects of dielectric disorder, which strongly affect the bare exciton dynamics in monolayer TMDCs. The low dephasing and weak effective inter-particle interaction enable the polaritons to travel across tens of micrometers with minimal energy dissipation, maintaining their partial coherence. These findings offer new insights into the dynamics of WS$_2$ polaritons at room temperature and the role of dielectric disorder in the TMDC systems strongly coupled to light. The demonstrated long-range ballistic flow and trapping of polaritons in the lowest energy states of a quasi-1D potential represent a significant step towards developing methods for manipulating and trapping polariton flow in TMDC-based polaritonic devices.

\section*{Methods}

{\bf Sample Fabrication.} A DBR chip splintered off a DBR substrate was placed on top of a polypropylene-carbonate (PPC) film \cite{Rupprecht2021}, which was initially spin-coated on top of a PDMS stamp supported by a glass slide. The two halfs of the SiO$_2$ $\lambda /2$-spacer were deposited by RF magnetron sputtering on top of the DBR chip and a DBR substrate, respectively, to ensure that the photonic field has its maximum at the centre of the microcavity. Further, a mechanically exfoliated monolayer WS$_2$ was transferred on top of the DBR substrate. Finally, the cavity was mechanically assembled at 130 $^{\circ}$C with a van der Waals stacking stage, at a temperatures at which the DBR chip detaches from the PPC film. 

{\bf Experimental Setup.} The photoluminescence spectra were measured with an in-house built optical setup, equipped with an array of lenses allowing for real-space (RS) and momentum-space (KS) imaging. The filtering in RS and KS were achieved with an edge-filter and an iris in the respective image planes. RS and KS imaging can be switched by flipping the lens, which images KS, in or out of the beam-path. The RS and KS spectra were measured with a spectrometer equipped with a CCD-camera and different spectrometers gratings, with 150 l/mm, 600 l/mm, and 1200 l/mm, allowing for energy resolutions down to $60~\mathrm{\mu eV/pixel}$. For the coherence measurements, we implemented a modified Michelson interferometer, where one arm is equipped with a retroreflector \cite{Kasprzak2006} that flips the image vertically. The output of the interferometer is fed onto the spectrometer and the interfering images are recorded using a CCD camera. The retroreflector arm is translated using a motorised stage to change the delay between the two arms.

{\bf Polariton Linewidth.} To estimate the polariton linewidth, we calculate the theoretical coherence times in an ideal system for the excitons and for the microcavity photons based on their Gaussian and Lorentzian linewidth contributions \cite{Reimer2016}:
\begin{equation}
    \tau_{X/C}^{H} = \left(\pi\Delta f^{H}_{X/C}\right)^{-1}, 
    \tau_{X/C}^{IH} = \sqrt{2ln(2)}\left(\sqrt{\pi} \Delta f^{IH}_{X/C}\right)^{-1}.
    \nonumber
\end{equation}
Here $\Delta f=\Delta E/h$. In the strong coupling regime, the polariton coherence time is determined by the exciton and cavity photon coherence times weighted by the excitonic Hopfield coefficient \cite{Deng2010}:
\begin{equation}
    \tau_{P}^{H/IH} (\left|X\right|^2) = \left(\left|X\right|^2/\tau^{H/IH}_X + (1-  \left|X\right|^2)/\tau^{H/IH}_C   \right)^{-1}.
    \nonumber
\end{equation}
By using the formulas above, we deduct the theoretical values for inhomogenous and homogenous broadening from the theoretical coherence times and obtain for the total linewidth of the resulting Voigt line profiles:
\begin{equation}
    \Delta f_P (\left|X\right|^2) = 0.5346 \Delta f_{P}^{H}+ \sqrt{0.2166 \Delta {f_{P}^{H}}^2 + \Delta {f_{P}^{IH}}^2}. \nonumber
\end{equation}
Without inhomogeneous broadening, the polariton linewidth can be directly calculated as:
\begin{equation}
\Delta E = |X|^2 \Delta E_X^X+(1-\Delta |X|^2) \Delta E_C,
\nonumber
\end{equation}
where $\Delta E_C$ is the cavity photon linewidth for the microcavity with the quality factor $Q\approx 3000$ (see Supplementary Information).

{\bf Interference Visibility.} 

The interference image measured by our camera $I_{tot}$ can be written as \cite{Askitopoulos2019}:
$$
I_{t}(r, \tau) = I_{o}(r) + I_{f}(r) + 2|g^{(1)}(r,\tau)|\sqrt{I_{o}(r)I_{f}(r)} \cos{(\kappa r + \phi)},
$$
where $\kappa$ and $\phi$ correspond to the fringe frequency and relative phase, respectively. The normalised interferograms presented in this work are calculated using the formula:
$$
V(r,-r,\tau) = \frac{I_{t}(r) - \left(I_{o}(r) + I_{f}(r)\right)}{2\sqrt{I_{o}(r)I_{f}(r)}}.
$$
The first order coherence function is the envelope of the normalised interferogram, as given by:
$$
|g^{(1)}(r,\tau)| \cos{(\kappa r + \phi)} = V(r,-r,\tau).
$$

\end{document}